\documentstyle[psfig]{mn}
\begin{document}

\title[Disc-jet coupling in 4U 1728--34]
{Disc-jet coupling in an atoll-type neutron star X-ray binary: 4U 1728--34 (GX
354--0)}
\author[S. Migliari et al.]
{S. Migliari$^1$\thanks{migliari@science.uva.nl},
R. P. Fender$^1$\thanks{rpf@science.uva.nl}, M. Rupen$^2$, 
P. G. Jonker$^3$, M. Klein-Wolt$^1$,  
\newauthor R. M. Hjellming$^2$, M. van der Klis$^1$\\ 
\\ 
$^1$ Astronomical Institute `Anton Pannekoek', University of Amsterdam,
and Center for High Energy Astrophysics, 
Kruislaan 403, \\
1098 SJ, Amsterdam, The Netherlands.\\
$^2$ National Radio Astronomy Observatory, Socorro, NM 87801, USA\\
$^3$Institute of Astronomy, Madingley Road, CB3 0HA, Cambridge \\
}

\maketitle

\begin{abstract}
We have analysed 12 simultaneous radio (VLA) and X-ray (RXTE) observations of
the atoll-type X-ray binary 4U 1728--34, performed in two blocks in 2000 and
2001.     
We have found that the strongest and most variable emission seems to be
associated with repeated transitions between hard (island) and softer (lower
banana) X-ray states, while weaker, persistent radio emission is observed
when the source is steadily in the hard X-ray state.      
There is a significant positive ranking correlation between the
radio flux density at 8.46~GHz and the 2--10~keV X-ray flux. 
Moreover, significant positive ranking correlations between radio flux density
and X-ray timing features (i.e. break and low-frequency Lorentzian
frequencies) have been found. 
These correlations represent the first evidence for a coupling between disc
and jet in an atoll-type X-ray binary. Furthermore, drawing an analogy between
the hard (island) state and the low/hard state of black hole binaries, we
confirm previous findings that accreting neutron stars are a factor of $\sim
30$ less `radio loud' than black holes. 

\end{abstract}


\section{Introduction}

In black hole candidates (BHCs) and in Z-type neutron star (NS) X-ray binaries
connections between inflow (disc) and outflow (jet) 
have been established. 
Many works (e.g. Hjellming \& Han 1995 and references therein; Falcke \&
Biermann 1996; Dhawan, Mirabel \& Rodr\'\i{}guez 2000; Fomalont, Geldzahler \&
Bradshaw 2001) suggest that all the radio emission from such systems
(including weak spatially unresolved emission) originates in jet-like
outflows. 
In persistent BHCs steady jet outflows 
are associated with the `low/hard' X-ray state, while not detected in the
`high/soft' state (Fender et al. 1999; Fender 2001a,b). In Z-type NSs
the radio emission seems to be strongest in the `horizontal' branch and
weakest in the `flaring' branch (Penninx et al. 1988; Hjellming \& Han 1995
and references therein). 
Atoll-type NS X-ray binaries share many X-ray spectral and timing
properties with BHCs (especially in the low/hard state; van der Klis 1994).  
However, only a few atolls are detected in radio band because of their
lower radio luminosity (Fender \& Hendry 2000). Hence, although they represent
the largest class of X-ray binaries ($\sim 45$\%, using
the `broader' definition of Fender \& Hendry 2000) no information on a possible
radio:X-ray coupling has been available until now.        


\begin{table*}
\centering
\caption{Modified Julian day (MJD), 2--10~keV (F$_{(2-10)}$)
and 2--60~keV (F$_{(2-60)}$) unabsorbed X-ray flux , 8.46~GHz (F$_{\rm 8.46}$)
and 4.86~GHz (F$_{\rm 4.86}$) radio flux densities (the upper limits are
$3\sigma$), break, L$_{h}$, L$_{b}$ and  L$_{u}$ frequency 
of 13 VLA 
observations of 4U 1728--34, 12 simultaneous with RXTE. The letters in the
first column refer to Fig.~1.}        
\label{tabspecB}
\vspace{0.2cm}
\begin{tabular}{l l l l l l l l l l}
\hline
\hline
&MJD & F$_{(2-60)}$$\times10^{-9}$& F$_{(2-10)}$$\times10^{-9}$  & F$_{\rm
8.46}$ &  F$_{\rm 4.86}$  & break  & L$_{b}$ &  L$_{h}$ & L$_{u}$\\ 
& &(erg s$^{-1}$cm$^{-2}$) &(erg s$^{-1}$cm$^{-2}$) & (mJy) & (mJy) & (Hz) &
(Hz) & (Hz) & (Hz)\\
\hline         
a&51638.53&$1.46\pm0.07$  &$1.03\pm0.05$ &$0.50\pm0.08$   & \ldots &$4.84\pm0.45$ &$15.36\pm0.64$&$28.59\pm1.63$ & \ldots\\   
b&51649.58$^{1}$&$0.88\pm0.10$ &$0.44\pm0.05$ &$<0.9$         & \ldots&$12.70\pm2.64$ &$21.95\pm0.68$&$51.49\pm3.55$ & $870\pm 3$\\ 
c&51669.51&$3.51\pm0.23$ &$2.25\pm0.15$ &$0.6\pm0.2$    & \ldots &$12.77\pm2.63$&$21.97\pm0.67$&$51.47\pm3.53$  & $870\pm4$\\ 
d&51677.32&$2.75\pm0.18$ &$1.54\pm0.10$ &$0.33\pm0.15$  & \ldots
&$4.85\pm0.45$ & $15.36\pm0.64 $ &$28.57\pm1.62$ & $563\pm23$\\ 
e&51685.47&$3.55\pm0.22$ &$1.81\pm0.11$ &$0.62\pm0.1$   & \ldots &$6.43\pm0.46$&$18.09\pm0.42$&$29.17\pm2.07$  & $610\pm13$\\ 
f&51695.34$^{1}$&$3.83\pm0.50$&$1.84\pm0.24$ &$<1.2$         & \ldots &$2.37\pm0.19$&\ldots&$15.40\pm0.58$ \ldots\\           
\hline
 &52056.37$^{2}$&\ldots  & \ldots                        &$0.18\pm0.02$&$0.19\pm0.05$& \ldots& \ldots & \ldots\\
g&52058.36$^{1}$&$4.66\pm0.31$&$2.42\pm0.16$&$0.11\pm0.02$&$<0.15$&$2.50\pm0.22$&
\ldots&$15.94\pm0.61$ & $616\pm19$\\ 

h&52061.35$^{1}$&$1.20\pm0.14$&$0.60\pm0.07$ &$0.09\pm0.02$&$<0.14$&$2.39\pm0.35$&\ldots&$13.75\pm1.13$ & \ldots \\ 
i&52063.31 & $1.31\pm0.13$&$0.61\pm0.06$ &$0.11\pm0.02$&$<0.13$&$1.29\pm0.18$&\ldots&$15.31\pm1.75$  & \ldots\\ 
j&52065.34&  $1.36\pm0.13$  &$0.62\pm0.06$ &$0.15\pm0.02$&$0.20\pm0.02$&$1.53\pm0.14$&\ldots&$10.98\pm0.59$ &\ldots\\
k&52067.30&  $1.43\pm0.25$  &$0.69\pm0.12$ &$0.16\pm0.02$&$<0.15$&$1.75\pm0.20$&\ldots&$12.19\pm0.74$  & \ldots\\ 
l&52069.29&  $1.48\pm0.11$ &$0.70\pm0.05$ &$0.09\pm0.02$&$<0.14$&$1.18\pm0.08$&\ldots&$9.83\pm0.41$ & $399\pm37$\\   
\end{tabular}
\flushleft 
{\bf 1:} the observation shows an X-ray burst;
{\bf 2:} not simultaneous with RXTE.
\end{table*}

4U~1728--34 (GX~354--0; Forman et al. 1976) is a low-mass X-ray binary and 
a type-I X-ray burster (Lewin 1976; Hoffman et al. 1976). 
From X-ray burst properties a distance to the source (accurate
to within 15\%: Kuulkers et al. 2002) of $5.2$~kpc, for a 1.4~M$_{\odot}$
mass NS, was obtained (Galloway et al. 2002; see also Basinska et al. 1984).
Hasinger \& van der Klis (1989) classified 4U~1728--34 as an atoll-type
X-ray binary.  
A multi-Lorentzian timing study of the power spectrum shows
well-defined and correlated features in both low and high
frequency range (for details see e.g. van Straaten
et al. 2002 and references therein).    
Timing properties are related to the position of the 
source in the colour-colour diagram (CD; e.g. M\'endez \& van der Klis 1999;
Di~Salvo et al. 2001; van Straaten et al. 2002), and therefore to the changing
mass accretion rate $\dot{\rm M}$ (which probably increases 
from the island to the banana state: e.g., Hasinger \& van der Klis 1989;
although there are secular changes on timescales of a few days or longer that
cause shifts in the CD: e.g. van der Klis 2001).
The continuum of the broadband 0.1--100~keV energy spectrum of the persistent
emission of 4U~1728--34 in the soft state is well fit by   
a soft thermal (blackbody or multitemperature disc blackbody) plus a
Comptonized component (Di~Salvo et al. 2000)

The optical counterpart of 4U~1728--34 cannot be detected due to the  high
extinction in the galactic center direction.  
In 1997 Mart\'\i{} et al. (1998) observed  4U~1728--34 in the radio
band (at 4.86~GHz) with the VLA, and after a few non-detections (with upper 
limits up to 0.32~mJy) succesfully detected its radio counterpart with a
variable flux density ranging between $\sim0.3$ and $\sim0.6$~mJy.      
They also detected a J and K-band infrared source (J=$19.6\pm0.4$ and
K=$15.1\pm0.2$) within one arcsec of the radio source.

\section{Observations and data analysis}

4U 1728--34 was observed on 13 occasions between 2000 and 2001 with the
VLA, 12 times simultaneously with RXTE observations. In Table~1 we show
the MJDs of the observations in 2000 ({\em a} to 
{\em f}) and in 2001 ({\em g} to {\em l} and the non-simultaneous with RXTE
at MJD 52056.37). The array was in configuration C 
during the observations in 2000, B during the non-simultaneous observation and
{\em g}, and CnB during {\em h}-to-{\em l}.              
Observation durations ranged from five min to a few hours;
flux densities measured at both 4.86 and 8.46 GHz are reported in Table
1. Flux calibration was achieved using J1331+305 or J0137+331; phase
calibration was performed using J1744--312 and data reduction using
AIPS. Absolute calibration of the flux density scale is estimated to
be accurate to $\sim 3$\%. No significant flux density variations were
found {\em within} each data set. Combining the data, the best-fit
coordinates for the radio counterpart are J2000 RA 17 31 57.687 +/-
0.013 Dec -33 50 01.11 +/- 0.18; this is consistent with the
position reported in Mart\'\i{} et al. (1998) within 1.5$\sigma$.
There is no evidence for any spatial extension during these observations.  

\begin{figure*}
\psfig{figure=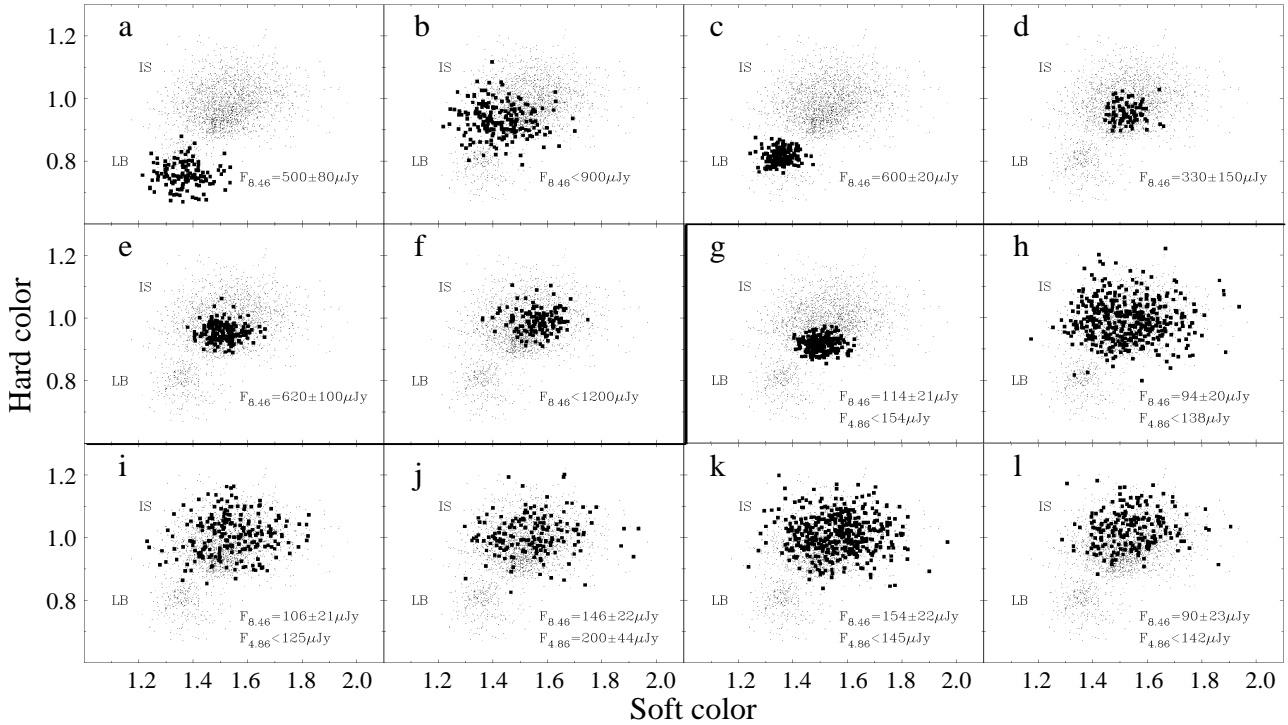,angle=0,width=17cm}
\caption{X-ray colour-colour diagrams (CDs) of the 12 RXTE/PCA observations of
4U~1728--34 simultaneous with VLA: Soft Colour=(3.5--6)~keV/(2--3.5)~keV, Hard
Colour=(9.7--16)~keV/(6--9.7)~keV. 
All the 12 simultaneous radio/X-ray observations are shown in each panel (the
small points represent 16 s os data). Marked with filled squares are the
individual observations in chronological order ({\em a}-to-{\em f} are the
observations in 2000 and {\em g}-to-{\em l} are the observations in 2001; see
Table~1).  
The corresponding radio flux densities in $\mu$Jy at 8.46~GHz and, when
available, at 4.86~GHz (upper limits are $3\sigma$) are also indicated.} 
\label{CD}
\end{figure*}

For the RXTE observations we have used data from the Proportional
Counter Array (PCA; for spectral and timing analysis) and the High Energy
X-ray Timing Experiment (HEXTE; only for spectral analysis). The durations of
the X-ray observations range between half an hour to a few hours.
We have used the PCA {\tt Standard2} data to produce the CD of
the RXTE observations (Fig.~\ref{CD}). The soft colour and the hard colour are
defined as the count rate ratio (3.5--6)~keV/(2--3.5)~keV and
(9.7--16)~keV/(6--9.7)~keV, respectively. We have normalised the colours of 4U
1728--34 to the colours of the Crab calculated with the closest observation
available to each 4U 1728--34 observation. The identification of the X-ray
states as island (IS) and lower banana (LB) are confirmed by timing
properties (see below; Di~Salvo et al. 2001; van Straaten et al. 2002).

For the spectral analysis of PCA {\tt Standard2} data  
we subtracted the background, estimated using {\tt pcabckest} v3.0, produced
the detector response matrix with {\tt pcarsp} v8.0 and
analysed the energy spectra in the range 3--20 keV. 
We extracted HEXTE energy spectra (channels 15--61) from both cluster A and B,
subtracted the background, corrected for deadtime using FTOOLS V5.2 and
analysed the spectra between 20 and 60 keV, except the observation {\em a} in
which the source was not detected above 40~keV. A systematic error of 0.75\%
was added to the PCA data. 
Four observations (see Table~1) show one X-ray burst in the PCA light curve. We
excluded the burst from the data and analysed only the spectra of the steady
persistent emission averaged over each observation.     
The spectra of {\em d}-to-{\em l} are well fit by an absorbed power law with a
high energy cutoff (in the range $\sim19-30$~keV) and a blackbody, plus a
$6.4-6.7$~keV Gaussian emission line. No high energy cutoff is necessary for
{\em a}, {\em b} and {\em c}. The equivalent   
hydrogen column density 
N$_{\rm H}$ was fixed to $2.5\times10^{22}$~cm$^{-2}$ (Hoffman et al. 1979;
Grindlay \& Hertz 1981; Foster et al. 1986; Di~Salvo et al. 2000). 

For the production of the power spectra we used {\tt event} data with a
time resolution of 125~$\mu$s. We rebinned the data 
in time to obtain a Nyquist frequency of 4096~Hz. For each observation we
created power spectra segments of 128~s length, cutting the first five energy
channels to avoid possible fake high frequency features of instrumental origin 
(Klein-Wolt, Homan \& van der Klis 2003 in prep.),           
and we removed X-ray bursts from the data, but no background and deadtime
corrections were performed. We averaged the power spectra and subtracted the
Poisson noise estimated between 3000 and 4000 Hz applying the standard method
by Zhang et al. (1995).  
We applied the Leahy et al. (1983) normalisation and then converted the power
spectra to squared fractional rms. For the fitting
procedures the multi-Lorentzian model was used in
the power times frequency representation (for details see
Belloni, Psaltis \& van der Klis 2001 and references therein).

\begin{figure}
\psfig{figure=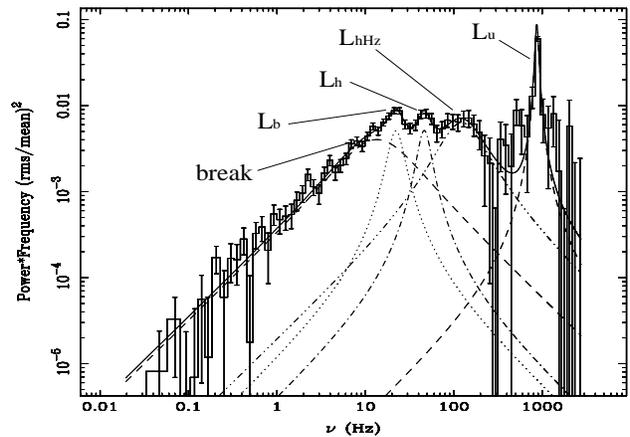,angle=0,width=8.2cm,height=5.7cm}
\caption{Power spectrum of the observation {\em c}. The break, 
L$_{b}$, L$_{h}$, L$_{hHz}$ and L$_{u}$ are shown.} 
\label{pow}
\end{figure}

The power spectra are fit using one broad Lorentzian to represent the low
frequency noise and the break frequency, one or two narrower
Lorentzians (L$_{b}$ and L$_{h}$: see van Straaten
van der Klis \& M\'endez 2003 for details on nomenclature), a broad Lorentzian
around 100~Hz (L$_{hHz}$) and narrow Lorentzians to fit the kHz QPOs (see
Fig.~\ref{pow} and Table~1).      
In five observations we have found only one kHz QPO.  
Based on the correlations of the frequency of the kHz QPOs with
their amplitude (Fig.~1 in M\'endez et al. 2000), and with the
frequencies of the other timing features (Fig.~3 in van Straaten et al. 2002),
we identified all of them as `upper' kHz QPOs (L$_{u}$).  
Both the `upper' and the `lower' kHz QPOs are found in {\em g}.  

\begin{figure}
\centering
\psfig{figure=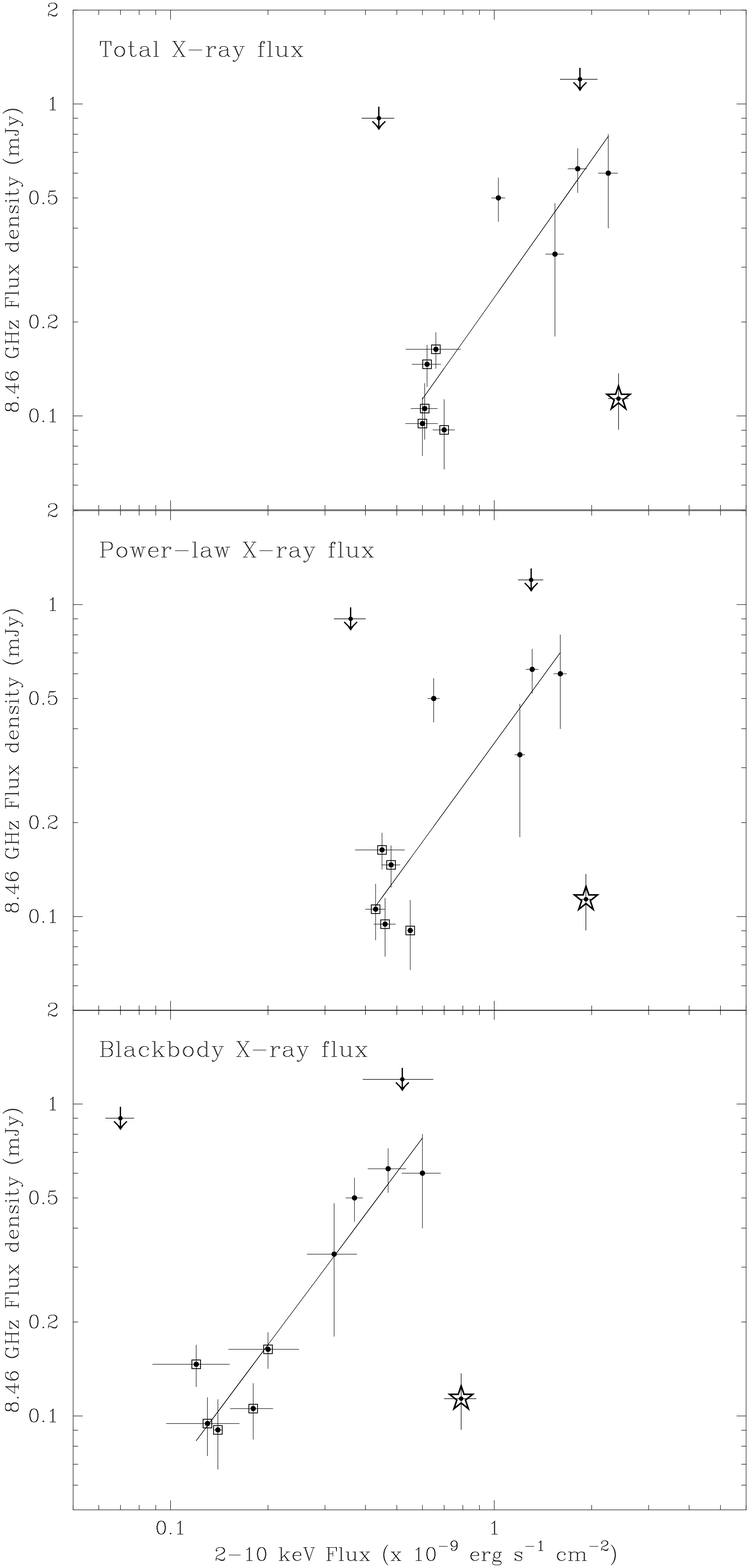,angle=0,width=7.5cm}
\caption{Radio flux density at 8.46~GHz versus the 2--10~keV unabsorbed
blackbody, power law and total (Blackbody+power law+Gaussian emission line)
flux. The line is the fit of the data excluding the radio flux density upper
limits (arrows) and the observation {\em g} (star) ($F_{R}\propto
F_{X}^{\Gamma}$, where $F_{R}$ is the radio flux density and 
$F_{X}$ is the X-ray flux) with $\Gamma=1.5\pm0.2$, $\Gamma=1.4\pm0.2$ and
$\Gamma=1.4\pm0.1$ for total, power law and blackbody 
X-ray flux respectively. The squares mark the observations in 2001.} 
 \label{fr-fx}
\end{figure}
\begin{figure}
\psfig{figure=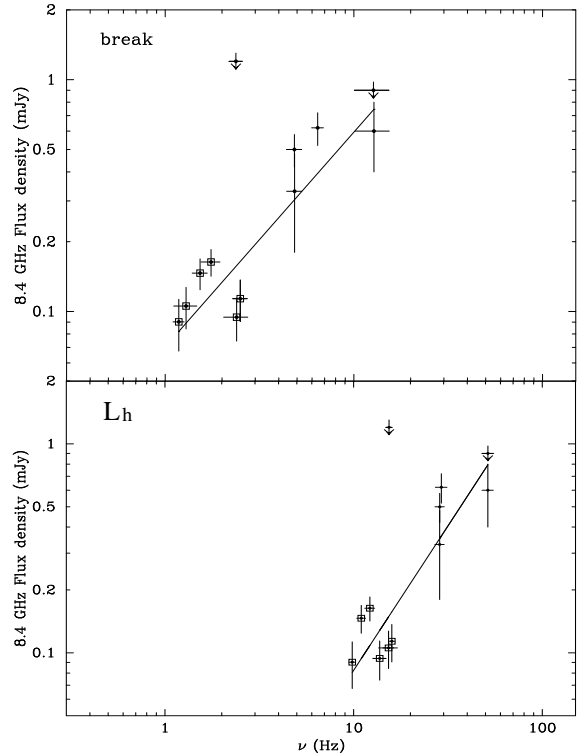,angle=0,width=7.5cm}
\caption{Correlations between break, L$_{h}$ frequencies 
and radio flux density at 8.46~GHz. The arrows are radio flux density
$3\sigma$ upper limits and the squares mark the observations in 2001.  
The lines are the fits of the correlations with a power law
($F_{R}\propto\nu^{\Gamma}$, where $F_{R}$ is the radio flux density and $\nu$
is the QPO frequency, with $\Gamma=0.9\pm0.1$ and $\Gamma=1.4\pm0.2$ for
break and L$_{h}$ respectively).}   
\label{F-br}
\end{figure}

\section{Overall pattern of behaviour}

Fig.~\ref{CD} shows the position in the CD of the 12 observations 
and the corresponding radio flux density (the upper limits are 3$\sigma$)
at 8.46~GHz and, where available, at 4.86~GHz. 
We see that the observations in 2000 (taken every $\sim10$
days) are mainly in the IS with two excursions to the LB. 
It seems that during this period the source was repeatedly transiting between
IS and LB. 
These observations show the highest radio flux density values up to 0.6
mJy with variations of $\sim0.3$~mJy between observations, and two
non-detections (both non-detections have $3\sigma$ limits above the other
measurements in IS, so are consistent with the other observations).   
In 2001 the observations (taken every 2--3~days) are steady in the IS, the
radio flux density is lower (around 0.1 mJy) and the variations are smaller
($\la0.06$~mJy) than in 2000.    
This indicates a possible association of radio flaring with 
transitions between hard (i.e. IS) and softer (i.e. LB)
X-ray states, and `quiescent/steady' radio emission with the hard (IS)
state.
We cannot find any measurable effect of type-I X-ray bursts on radio
emission, although we notice that the bursts are detected when the radio flux
seems to be low: in two observations of 2001 and in the two observations
of 2000 with no radio detections (see Table~1).   

The dual-frequency radio measurements are not good enough to seriously
constrain the radio spectrum (see Table~1). They are in most cases consistent
with both flat ($\alpha\sim0$, where $S_{\nu}\propto\nu^{\alpha}$ and
$S_{\nu}$ is the radio flux density at a certain frequency $\nu$) spectrum
radio emission as observed from low/hard state black holes,  
or the optically thin emission ($\alpha\sim -0.6$) observed from X-ray
transients (i.e. non-simultaneous observation: $\alpha=-0.14\pm0.36$;
{\em g}: $\alpha>-0.55$; {\em h}: $\alpha>-0.69$; {\em i}: $\alpha>-0.31$;
{\em j}: $\alpha=-0.56\pm0.14$; {\em k}: $\alpha>-0.21$; {\em l}:
$\alpha>-0.81$).     

\section{Radio:X-ray correlations}

Fig.~\ref{fr-fx} shows the radio flux density as a function of the soft
spectral component (i.e. blackbody), the hard spectral component (i.e. power
law) and the total unabsorbed 2--10~keV X-ray flux (this range is chosen to
allow a direct comparison with the radio:X-ray flux correlation in BHCs; e.g
Gallo Fender \& Pooley 2003; see \S~5) of the 12 observations. Excluding
the observation {\em g} (the star; we will discuss this point in \S~5), there
are significant positive ranking correlations between the radio flux density
and the X-ray fluxes ($99$\%, 97\% and $98$\% significance respectively with 
blackbody, power law and total flux; the fitting power-laws
$F_{R}\propto F_{X}^{\Gamma}$, where $F_{R}$ is the radio flux density and
$F_{X}$ is the X-ray flux, shown in Fig.~\ref{fr-fx}
have $\Gamma$ of $1.4\pm0.1$, $1.4\pm0.2$ and $1.5\pm0.2$ for blackbody,
power-law and total X-ray flux respectively).   
This indicates that the jet power is correlated to the accretion
rate as inferred from X-ray flux.  

In Fig.~\ref{F-br} we show the frequency of the break and L$_{h}$ as a
function of the radio flux density. There is 
a significant positive ranking correlation between break, L$_{h}$ frequencies
and radio flux density (99\% and 98\% significance respectively; in Fig.~4 the
fitting power laws $F_{R}\propto\nu^{\Gamma}$, where $F_{R}$ is the radio flux
density and $\nu$ is the frequency, have $\Gamma=0.9\pm0.1$ and
$\Gamma=1.4\pm0.2$ for break and L$_{h}$ respectively).    
There is also a hint (with only four points, treating
with caution the observation {\em g}; see also \S~5) of a correlation between
L$_{u}$ frequency and the radio flux density.       
An opposite (qualitative) behaviour was found by Muno et al. (2001) for
the BHC GRS~1915+105; they found that the `radio plateau' observations
(i.e. radio flux densities $\sim 100$~mJy at 15.2~GHz) show lower 0.5--10~Hz
QPO frequencies than `radio faint' observations (i.e. radio flux densities
$\la$ 20~mJy at 15.2~GHz). 

The QPO frequencies are generally interpreted as being
related to the motion of matter in the accretion disc at a certain
radius. In particular, in low magnetic field NS systems, the kHz QPOs are
thought to be related to motion of matter a few stellar radii from the star
(see van der Klis 2000 for a review), and, as in e.g. Miller,
Lamb and Psaltis (1998), correspond to the inner radius of the Keplerian disc.
According to most models (see e.g. van der Klis 2001 and references therein)
the kHz QPOs (and also other timing features like the
break and L$_{h}$ that generally correlate with them; Di~Salvo et al. 2000;
M\'endez, van der Klis \& Ford 2001; van Straaten et al. 2002) are related to
the disc mass accretion rate.  
Therefore our radio flux/X-ray timing correlations represent independent
(i.e. different from radio:X-ray flux correlations) evidence for a coupling
between accretion and outflow rates. 
As a caveat, we note that the correlations discussed above are dominated
by the difference between the two blocks of data, in 2000 and 2001
respectively. While the correlations themselves are not in doubt, further
observations are required to establish if there is a smooth relation
between the blocks or a more bimodal form of behaviour.

\section{Discussion}

We know that the X-ray flux does not trivially trace the disc mass accretion
rate; `parallel tracks' are observed between kHz QPO frequencies 
and X-ray luminosity (e.g. M\'endez et al. 1999; Ford et al. 2000; M\'endez et
al. 2001).  
The strong coupling between spectral and timing properties in
X-ray binaries, suggests that the QPOs are actually a more straightforward
indicator, rather than luminosity, if not of the absolute value of $\dot{\rm
M}$ at least of variations of the disc mass accretion rate and maybe
of the inner radius of the accretion disc (see van der Klis 2001).             
Therefore, the radio:X-ray flux and even more the radio flux/X-ray timing
correlations translate into the first evidence for a coupling between 
the accretion disc inflow and the jet outflow in an atoll source. 

4U 1728--34 shows a correlation between radio flux density and X-ray
flux qualitatively similar to that found for BHCs in the low/hard state
(Hannikainen et al. 1998; Corbel et al. 2000; Corbel et al. 2002; Gallo et
al. 2003), although the index $\Gamma$ of the coupling (see Fig.~3) is
rather steeper. 
The observation {\em g} (star in Fig.~\ref{fr-fx}, has the highest X-ray flux
in our sample and a low (compared to what `expected' from the radio:X-ray flux
correlation we have found for the other observations) radio flux
density. Since {\em g} shows the same low frequency timing properties of the
other observations in 2001, this high X-ray flux can be interpreted as due to
a `parallel track' effect (see e.g. M\'endez et al. 2001).  
Furthermore, in the power spectrum, although the low frequency
features are in agreement with previous observations at the same position in
the CD, the relative strengths of the kHz QPOs we find (the `lower' kHz QPO is
stronger than the `upper') are usually observed in softer observations,
i.e. in the banana state rather than in the IS (e.g. van Straaten et
al. 2000).   
A comparison with BHCs (see e.g. Gallo et al. 2003) also show that {\em g} is
almost at the same X-ray luminosity (using a distance of 5.2~kpc: Galloway et
al. 2002) as the `radio quenching' in BHCs; this suggests (nothing more than 
this, since it is only one point) that suppression (`quenching') of the radio
jet may occur above some luminosity, as in BHCs.  
It is interesting to note that 4U 1728--34, at the same X-ray luminosity as
e.g. Cyg X-1 (Gallo et al. 2003), show a radio luminosity $\sim30$ times less
than Cyg X-1 (i.e. scaled to 1 kpc the radio flux densities are $F_{4U
1728-34}\sim 2.5$~mJy and $F_{Cyg X-1}\sim 75$~mJy). This confirms (also
quantitively) the Fender \& Kuulkers (2000) finding of radio `loudness'
difference between atoll-type NSs and BHCs.               

X-ray (i.e. mainly accretion) properties in atoll sources and in BHCs (in
low/hard state) seem to be qualitatively the same (e.g. van der Klis
1994). This suggests that the same physical processes take place in both type
of sources. What about the physical processes that connect inflow (disc)
and outflow (jet) matter in X-ray binaries? The key to answering this question
lies in the study of atoll vs BHCs disc-jet coupling. 
The results presented in this paper may be a first step in that direction.  


\section*{Acknowledgements}
The National Radio Astronomy Observatory is a facility of the National
Science Foundation operated under cooperative agreement by Associated
Universities, Inc. We would like to thank Steve van Straaten and Tiziana Di
Salvo for useful discussions.

\end{document}